\documentclass[12pt]{article}
\usepackage{amssymb,epsfig,setspace}
\begin{document}

\title{
A superconvergent representation of the Gersten-Nitzan and Ford-Webber
nonradiative rates
}

\author{
Alexander Moroz\thanks{http://www.wave-scattering.com}
\\
Wave-scattering.com
} 

\date{}

\maketitle

\begin{center}
{\large\sc abstract}
\end{center}
An alternative representation of the quasistatic nonradiative 
rates of Gersten and Nitzan [J. Chem. Phys. 1981, 75, 1139] 
and Ford and Weber [Phys. Rep. 1984, 113, 195] 
is derived for the respective parallel and perpendicular dipole orientations.
Given the distance $d$ of a dipole from a sphere surface of radius $a$,
the representations comprise four elementary analytic functions 
and a modified multipole series taking into account 
residual multipole contributions. The analytic functions could be
arranged hierarchically according to decreasing singularity at the short 
distance limit $d\rightarrow 0$, ranging from $d^{-3}$ over $d^{-1}$ to $\ln (d/a)$. 
The alternative representations
exhibit drastically improved convergence properties.
On keeping mere residual dipole contribution of the modified multipole series, the  
representations agree with the converged rates on at least   
$99.9\%$ for all distances, arbitrary particle sizes and emission wavelengths, and 
for a broad range of dielectric constants.
The analytic terms of the representations reveal a complex distance dependence and
could be used to interpolate between the familiar $d^{-3}$ 
short-distance and $d^{-6}$ long-distance behaviors with an unprecedented accuracy. 
Therefore, the representations could be especially useful for 
the qualitative and quantitative understanding 
of the distance behavior of nonradiative rates 
of fluorophores and semiconductor
quantum dots involving nanometal surface energy transfer in the presence 
of metallic nanoparticles or nanoantennas.
As a byproduct, a complete short-distance asymptotic 
of the quasistatic nonradiative rates is derived.
The above results for the nonradiative rates translate
straightforwardly to the so-called image enhancement factors $\Delta$,
which are of relevance for the surface-enhanced Raman scattering.

\vspace*{1.9cm}

\newpage

\section{Introduction}
The distance dependence of the nonradiative rate  
of a fluorophore or a semiconductor
quantum dot relative to a metallic nanoparticle (MNP), or 
a nanoantenna, is of crucial importance for a number of novel techniques
and nanodevices combining biomolecules, quantum dots, and MNP 
\cite{DCL,MTN,Dulk,Lak5,YJJ,JSS,SRW,PMS,HSP}.
The MNPs can quench fluorescence as much as $100$ times better than 
other quenchers of fluorescence, such as
DABCYL, and open new perspectives in the use of hybrid materials 
as sensitive probes in fluorescence-based detection assays \cite{DCL}.
In the case of single-stranded (ss) DNA
probes labeled with a thiol at one end and a dye at the other,
the ssDNA molecules self-organize into a constrained
conformation on the MNP surface and the fluorophore
is completely quenched by the particle through a distance-dependent process.  
Upon target (complementary oligo) binding, the constrained 
DNA conformation is opened because of a dramatic 
increase in the DNA rigidity that becomes double-stranded 
after hybridization. Thereby the fluorophore is separated 
from the particle surface, and the fluorescence 
of the hybrid molecule can increase by a 
factor of as much as several thousand as it binds to a 
complementary ssDNA \cite{DCL}.
This structural change generates a fluorescence signal that is highly sensitive and
specific to the target DNA \cite{DCL,MTN}, resulting in a molecular beacon
that can detect
minute amounts of oligonucleotide target sequences in a pool of
random sequences and provide an
improved detection of a single mismatch in a competitive
hybridization assay for DNA mismatch detection \cite{DCL}.
MNP acceptors can also be employed for probing changes in distances for 
protein interactions on DNA using a molecular ruler approach 
involving nanometal surface energy transfer (NSET) from optically 
excited organic fluorophores to small MNPs \cite{YJJ,JSS}.
Whereas the traditional F\"{o}rster resonance energy transfer (FRET) 
involving molecular acceptors and donors is efficient for the
acceptor-donor separations up to $100$ \AA, the use of MNP acceptors 
could more than double the traditional F\"{o}rster 
range up to $220$ \AA\, \cite{YJJ,JSS}. As it becomes 
apparent below, the reason for the extension of 
the traditional F\"{o}rster range is a relaxed distance dependence 
relative to FRET, because the particle-fluorophore interaction 
could no longer be described by sole dipole-dipole interaction.

\begin{figure}[tbp]
\begin{center}
\epsfig{file=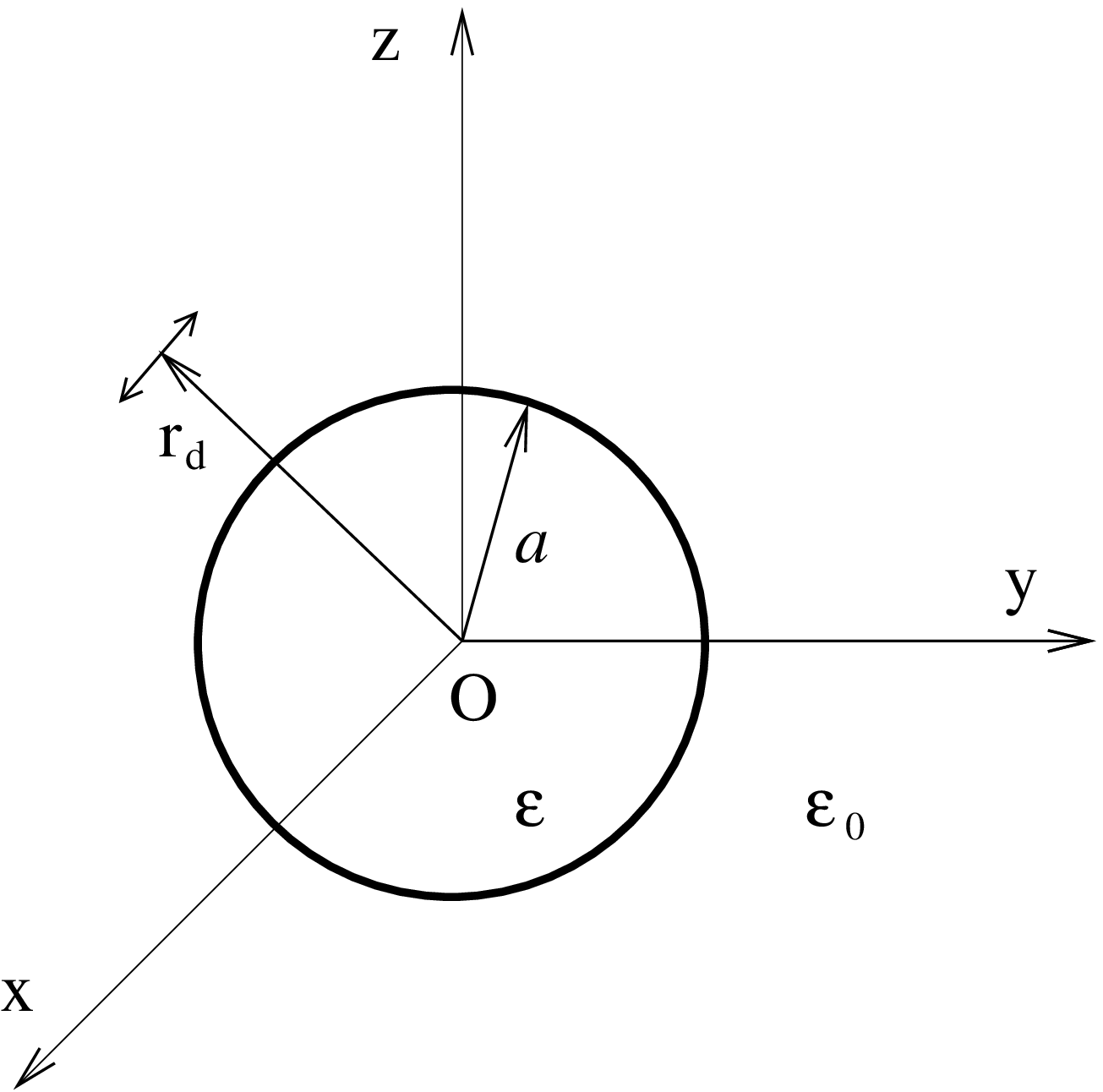,width=14cm,clip=0,angle=0}
\end{center}
\caption{}
\label{dpsg}
\end{figure}

The nonradiative rates are derived from the 
total Ohmic loss, or the Joule heating, which is in general 
determined as the power $P_{abs}$ absorbed inside the particle 
\cite{Rup,KLG2,GN,FoW1,AMap,AMcl,MKP,AMoc}
\begin{equation}
P_{abs} =\int_V Q({\bf r}) \, d{\bf r}
\label{pnrad}
\end{equation}
where the volume integral extends over entire absorbing region. 
The quantity $Q$ is the {\em steady} (averaged) inflow of energy $Q$ per 
unit time and unit volume from the external sources which maintain the field,
and which in the gauss units is given by
\begin{equation}
Q=\frac{c k_0}{8\pi} \left(\varepsilon'' |{\bf E}|^2 + \mu'' |{\bf H}|^2 \right)
\label{endis}
\end{equation}
where $k_0$ is the vacuum wave vector and $\varepsilon''$ ($\mu''$) 
is the imaginary part of the dielectric function 
(magnetic permeability) at the observation point. 
Alternatively, in the special case of an 
oscillating dipole source ${\bf p} e^{-i\omega t}$,
one could consider the dipole moving in a local
field produced by its surface image and calculate the dipole dissipated 
power by (see eq 3.1 of ref \cite{FoW1})
\begin{equation}
P_{abs}=\frac{\omega}{2}\, \mbox{Im } ({\bf p}^*\cdot {\bf E}_{ind})
\label{fwf}
\end{equation} 
where ${\bf E}_{ind}e^{-i\omega t}$ is the electric field induced 
by the dipole at the dipole position ${\bf r}_d$, i.e. by the current source
\begin{equation}
{\bf j}({\bf r},t) = -i\omega{\bf p}  e^{-i\omega t}\delta({\bf r}-{\bf r}_d)
\end{equation}
The nonradiative rate $\Gamma_{nr}$ is then given by the 
correspondence principle 
\begin{equation}
\Gamma_{nr} =\frac{P_{abs}}{\hbar\omega}
\label{qmcm}
\end{equation}
For the sake of definiteness, let the sphere center be
located at the coordinate origin, the
dipole position denoted be described by 
the position vector ${\bf r}_d$,
$r_d=|{\bf r}_d|>a$ (see Figure \ref{dpsg}), and 
the respective $\varepsilon$ and $\varepsilon_0$ be 
the sphere and host dielectric constants.
Although exact expressions for the total, radiative, 
and nonradiative rates for a dipole interacting with a spherical particle 
\cite{Rup,KLG2,Ch,Chew}, including the case of a multi-coated sphere 
\cite{AMap,AMcl}, are known, and the F77 source codes could be freely 
downloaded \cite{AMcd}, the {\em quasi-static} results 
of Gersten and Nitzan \cite{GN,GNa} and Ford and Weber \cite{FoW1}
(cf eqs \ref{gmnrperp} and \ref{gmnrpar} below) have 
nevertheless still retained their value. 
The {\em quasi-static} results for small 
enough sphere radius $a$ agree rather well with exact
calculations \cite{AMap,MKP,AMoc} and experiment
\cite{SRW}, and allow one for a valuable qualitative
analytical insight. For a small separation $d$ of a fluorophore from
a sphere surface, such that the dimensionless distance parameter $\delta=d/a\ll 1$,
the quasi-static results predict a $d^{-3}$ dependence of 
the nonradiative rates (see eqs \ref{gmnrperpl} and \ref{gmnrparl} below).
The $d^{-3}$ short-distance dependence corresponds exactly to 
the case of a dipole located at the distance
$d$ above a half-space characterized by the dielectric constant $\varepsilon$ 
\cite{GN,FoW1,GNa}. However, as shown in Figures \ref{gnnra10l612}, \ref{asmpl340a100}
the $d^{-3}$ short-distance dependence begins 
to deviate rapidly from the converged rates for $\delta\gtrsim 0.1$ with increasing 
$\delta$. Thus, for a small
MNP with $a\lesssim 10$ nm, the $d^{-3}$ short-distance dependence 
of the quasi-static nonradiative rates turns out to be valid only
in an immediate proximity to the MNP ($d \lesssim 1$ nm)
where the very use of the quasi-static rates is questionable.
In the present work we remedy the above shortcomings by providing 
an analytic formula, which describes the distance behavior of the
nonradiative rates over all distances with an unprecedented accuracy.

In what follows, we derive in Section \ref{sc:gn}
an alternative superconvergent representation of the 
quasistatic nonradiative rates (eqs \ref{gmnraf1}, 
\ref{gmnraf2}, \ref{mms} below)
with drastically improved convergence properties. 
The superconvergent representation will be obtained 
by dividing the total contribution of each given multipole 
into a number of partial contributions (eqs \ref{opex1}, \ref{opex2} below) so that 
the essential part of the partial contributions could 
be summed up analytically (eqs \ref{sprdf}, \ref{spardf} below) 
leaving behind a residual infinite multipole series.
The first four exact sums are expressed in terms of elementary analytic
functions of $u=(a/r_d)^2$, with each subsequent function
having a lesser singularity at the short 
distance limit $d\rightarrow 0$, ranging from 
$d^{-3}$ over $d^{-1}$ to $\ln (d/a)$.
Convergence properties of the alternative 
representations are examined in Section \ref{sc:cnv}.
On keeping mere residual dipole term of the modified multipole series, the  
alternative representations are demonstrated to agree with the converged 
quasistatic nonradiative rates (eqs \ref{gmnrperp} and \ref{gmnrpar}) 
of Gersten and Nitzan \cite{GN,GNa} and Ford and Weber \cite{FoW1} on at least   
$99.95\%$ over entire length interval, 
for arbitrary particle sizes and emission wavelengths, and 
for a broad range of dielectric constants.
In Section \ref{sc:at} the contribution of the analytic terms relative 
to that of the residual multipole series is analyzed.
The alternative representations make it also possible 
to derive a complete short distance asymptotic of the classic quasi-static 
nonradiative rates including all singular terms in 
the distance parameter $\delta$. This is performed in Section \ref{sc:asp}.
A two-term asymptotic will be shown  to provide 
a significant improvement over the conventional 
$d^{-3}$ asymptotic over extended length scale.  
Keeping more terms in the asymptotic expansion improves the asymptotic
in an immediate sphere proximity $\delta\ll 0.1$, but it comes at the expense
of worsening the precision for $\delta\gtrsim 0.25$.
In Section \ref{sc:d3t} common approaches for fitting experimental
nonradiative rates  are critically examined
in light of the results of the present paper. 
In Section \ref{sec:ief}, the results for the nonradiative rates 
are shown to straightforwardly translate 
to the so-called image enhancement factors $\Delta$, which are of relevance 
for the surface-enhanced Raman scattering (SERS). 
The alternative superconvergent representations provide a 
significant improvement over 
the familiar $d^{-3}$ short-distance and $d^{-6}$ long-distance behaviors, 
and could be especially useful for the
qualitative and quantitative understanding of the 
distance behavior of nonradiative rates
of fluorophores and semiconductor
quantum dots involving nanometal surface energy transfer in the presence 
of metallic nanoparticles or nanoantennas \cite{DCL,MTN,Dulk,YJJ,JSS}.
The remaining points are discussed in Section \ref{sec:disc}. 
We then conclude with Section \ref{sec:conc}.

\section{Superconvergent representations of the quasistatic 
nonradiative rates}
\label{sc:gn}
When calculating decay rates for a dipole 
interacting with a spherical particle, one 
distinguishes the cases of a perpendicular ($\perp$) and
parallel ($\parallel$) dipole orientation relative to the sphere surface.
The case of a general dipole orientation is then reduced to a linear
combination of the two particular cases.
Let $\bar{\varepsilon}=\varepsilon/\varepsilon_0$ 
be the relative dielectric constant, and let 
$\mbox{\boldmath $\mu$}$ denote the molecular dipole.
The {\em nonradiative rates} of Gersten and Nitzan \cite{GN,GNa} 
for a {\em perpendicular} dipole orientation are
(cf eq B.24' of ref \cite{GNa})
\begin{eqnarray}
\Gamma_{nr} (\perp) &=& - \frac{|\mbox{\boldmath $\mu$}|^2}{2\hbar \varepsilon_0 a^3} 
\sum_{l=1}^\infty    (2l+1) \frac{(l+1)^2}{l} 
\left(\frac{a}{r_d}\right)^{2l+4} \mbox{Im }\frac{1}{\bar{\varepsilon}+ \frac{l+1}{l}}
\nonumber\\
&=&  \frac{|\mbox{\boldmath $\mu$}|^2}{2 \hbar \varepsilon_0 a^3} 
\sum_{l=1}^\infty   (l+1)^2
\left(\frac{a}{r_d}\right)^{2l+4} 
   \mbox{Im }\frac{\bar{\varepsilon} - 1}{\bar{\varepsilon}+ \frac{l+1}{l}}
\label{gmnrperp}
\end{eqnarray}
whereas the rates for a {\em parallel} dipole orientation are
(see eq B.45' of ref \cite{GNa})
\begin{eqnarray}
\Gamma_{nr}(\parallel ) &=& - 
   \frac{|\mbox{\boldmath $\mu$}|^2}{4 \hbar \varepsilon_0 a^3} 
      \sum_{l=1}^\infty  (l+1)(2l+1) \left(\frac{a}{r_d}\right)^{2l+4}
         \mbox{Im }\frac{1}{\bar{\varepsilon}+ \frac{l+1}{l}}
\nonumber\\
&=& \frac{|\mbox{\boldmath $\mu$}|^2}{4\hbar \varepsilon_0 a^3} 
\sum_{l=1}^\infty  l(l+1) \left(\frac{a}{r_d}\right)^{2l+4}
\mbox{Im }\frac{\bar{\varepsilon} - 1}{\bar{\varepsilon}+ \frac{l+1}{l}}
\label{gmnrpar}
\end{eqnarray}
where Im denotes the imaginary part.
The respective 2nd lines of eqs \ref{gmnrperp} and \ref{gmnrpar}  
have been obtained on using 
\begin{equation}
\mbox{Im }\frac{1}{\bar{\varepsilon}+ \frac{l+1}{l}} = -\frac{l}{2l+1} 
\, \mbox{Im }\frac{\bar{\varepsilon} - 1}{\bar{\varepsilon}+ \frac{l+1}{l}}
\label{imid}
\end{equation}
which follows from the identity:
\begin{equation}
\bar{\varepsilon} - 1 = \bar{\varepsilon} + \frac{l+1}{l} - \frac{2l+1}{l}
\end{equation}
Occasionally, the Gersten and Nitzan \cite{GN,GNa} expressions are
written with an {\em intrinsic} molecular dipole moment 
$\mbox{\boldmath $\mu$}_0$ that
is related to $\mbox{\boldmath $\mu$}$ by (cf. eqs B.13, B.15 of ref \cite{GN})
\begin{equation}
\mbox{\boldmath $\mu$}=\mbox{\boldmath $\mu$}_0 + \alpha E_{loc}
 = \frac{\mbox{\boldmath $\mu$}_0}{1-\Delta}
\label{ief}
\end{equation}
Here $E_{loc}$ is the local electric field that is in general
affected by the presence of a sphere, $\alpha$ is 
the molecule polarizability, and $\Delta$ is 
a corresponding {\em image enhancement factor}. 
The {\em intrinsic} 
molecular dipole moment $\mbox{\boldmath $\mu$}_0$ is what would be found for 
a totally isolated molecule far away the sphere.
$\Delta$ is significant only very close to the surface,
in a region where the use of a classical approach 
and of the point dipole model for the molecule are questionable.
Hence $\Delta$ may be disregarded in computing the actual decay rates \cite{GNa}.
The formulas (eqs \ref{gmnrperp} and \ref{gmnrpar}) were independently
confirmed by Ford and Weber, who determined 
the power dissipated by an arbitrary oriented dipole as (eq 3.47 of ref \cite{FoW1})
\begin{equation}
P_{abs} = \frac{\omega}{2 \varepsilon_0 a^3}\, \sum_{l=1}^\infty  (l+1)
\left[(l+1)\mbox{\boldmath $\mu$}_\perp^2 +\frac{l}{2}\mbox{\boldmath $\mu$}_\parallel^2\right]
\left(\frac{a}{r_d}\right)^{2l+4}
    \mbox{Im }\frac{l (\varepsilon -\varepsilon_0)}{l \varepsilon + (l+1)\varepsilon_0}
\label{fwdp}
\end{equation}
where $\mbox{\boldmath $\mu$}_\perp$ and $\mbox{\boldmath $\mu$}_\parallel$
are the respective perpendicular and parallel components of 
$\mbox{\boldmath $\mu$}=\mbox{\boldmath $\mu$}_\perp+\mbox{\boldmath $\mu$}_\parallel$. 
When substituted in the correspondence principle (eq \ref{qmcm}), 
the result for $P_{abs}$ leads to the nonradiative rates 
that are {\em identical} to those 
obtained by Gersten and Nitzan \cite{GN,GNa}.

In what follows, we employ a simple recipe 
to arrive at an alternative representation  
of the Gersten and Nitzan (GN) expressions (eqs \ref{gmnrperp} 
and \ref{gmnrpar}) with
drastically improved convergence properties. First note that
the `Im' sign in eqs \ref{gmnrperp} and \ref{gmnrpar}
can be brought in front of the summation sign.
Afterward, the task of deriving an alternative representation 
of the power series in eq \ref{gmnrperp} for a {\em perpendicular} 
dipole orientation  
reduces effectively to that for the power series
\begin{equation}
S_\perp = \sum_{l=1}^\infty  (l+1)^2 
        \frac{\bar{\varepsilon} - 1}{\bar{\varepsilon}+ 1 + \frac{1}{l}}
           \left(\frac{a}{r_d}\right)^{2l}
=
  \frac{\bar{\varepsilon}- 1}{\bar{\varepsilon} + 1}
          \sum_{l=1}^\infty  \frac{l(l+1)^2}{l + v} u^{l}
\label{sprd}
\end{equation}
where $u$ and $v$ are the shorthands for
\begin{eqnarray}
u = \left(\frac{a}{r_d}\right)^2, 
          && v = [\bar{\varepsilon} +1]^{-1}=\frac{\varepsilon_0}{\varepsilon+\varepsilon_0}
\label{uvd}
\end{eqnarray}
In terms of $S_\perp$,
\begin{equation}
\Gamma_{nr} (\perp) =  
  \frac{|\mbox{\boldmath $\mu$}|^2}{2\hbar\varepsilon_0 a^3} \, u^2 \, \mbox{Im } S_\perp
\label{gmnrperps}
\end{equation}
Now follows a point of crucial importance. This is an iterative decomposition
of the coefficients of the power series in eq \ref{sprd} into decreasing powers
of $l$. The iterative decomposition could be illustrated as follows.
We begin to single out the leading $l^2$ dependence as follows
\begin{equation}
\frac{l(l+1)^2}{l + v} = \frac{(l+v-v)(l+1)^2}{l + v} 
           = (l+1)^2   - \frac{v (l+1)^2}{l + v} 
\label{ist1}
\end{equation}
Note that the last term in eq \ref{ist1} is effectively of the order $l$
for $l\gg 1$, which is smaller than the $l^2$ order of the initial term. 
One can thus repeat the preceding step on the 
last term in eq \ref{ist1}, etc. By continuing the 
above iterative decomposition, one arrives at
\begin{eqnarray}
\frac{l(l+1)^2}{l+v} &=& 
   (l+1)^2 -v(l+1) - v (1-v) - \frac{v(1-v)^2}{l} + \frac{v^2(1-v)^2}{l^2}
                       - \frac{v^3(1-v)^2}{l^2(l+v)}
\nonumber\\
 &=& 
   l^2 + l(2-v) + (1-v)^2 - \frac{v(1-v)^2}{l} + \frac{v^2(1-v)^2}{l^2}
                       - \frac{v^3(1-v)^2}{l^2(l+v)}
\label{opex1}
\end{eqnarray}
On applying the above decomposition in eq  \ref{sprd}, 
the total contribution of each given multipole is divided into a number of 
partial contributions. The advantage of the decomposition is in that
the first four partial contributions in eq. \ref{opex1} 
could easily be summed up analytically. Indeed, upon substituting 
eq \ref{opex1} back into eq \ref{sprd}, one obtains on using
elementary summation formulas (eqs \ref{sf1}- \ref{sf4} of Appendix \ref{sec:sf})
\begin{eqnarray}
\lefteqn{
u^2 S_\perp = \frac{\bar{\varepsilon}- 1}{\bar{\varepsilon} + 1} \, u^2\,
\left[
\frac{u(1+u)}{(1-u)^3} + \frac{u(2-v)}{(1-u)^2} 
\right.
}
\nonumber\\
&&
\left.  + \frac{u(1-v)^2}{1-u} + v(1-v)^2\, \ln(1-u) + v^2(1-v)^2\, F(u,v) \right]
\label{sprdf}
\end{eqnarray}
where the remainder
\begin{equation}
F(u,v)= \sum_{l=1}^\infty \frac{u^l}{l(l+v)} 
     = \sum_{l=1}^\infty \frac{u^l}{l^2}  - v\, 
                 \sum_{l=1}^\infty \frac{u^l}{l^2(l+v)}
\label{fuv}
\end{equation}
takes into account a {\em residual contribution} of the original multipoles that
has not been comprised by the analytic functions in the square bracket of 
eq \ref{sprdf}. On substituting eq \ref{sprdf} in eq \ref{gmnrperps}, 
\begin{eqnarray}
\lefteqn{
\Gamma_{nr} (\perp) =  
  \frac{|\mbox{\boldmath $\mu$}|^2}{2\hbar\varepsilon_0 a^3} \,
\mbox{Im }\left\{ \frac{\varepsilon - \varepsilon_0}{\varepsilon + \varepsilon_0} \,
\left[
\frac{u^3(1+u)}{(1-u)^3}  + \frac{u^3}{(1-u)^2} 
\frac{2 \varepsilon+ \varepsilon_0}{\varepsilon+ \varepsilon_0}
+  \frac{u^3}{1-u}  \frac{\varepsilon^2}{(\varepsilon+ \varepsilon_0)^2}
\right.\right.
}
\nonumber\\
&&
\left.\left.
+ \frac{\varepsilon_0\varepsilon^2}{(\varepsilon+ \varepsilon_0)^3}\,  
                    u^2\, \ln(1-u)  
  + \frac{\varepsilon_0^2\varepsilon^2}{(\varepsilon +                                   \varepsilon_0)^4}\,  u^2\, F(u,v)
\right]\right\}
\hspace*{2cm}
\label{gmnraf1}
\end{eqnarray}
where we have substituted 
for $v$ in eq \ref{sprdf} according to eq \ref{uvd}. 

The task of arriving at an alternative representation  
of the power series in eq \ref{gmnrpar}
for a {\em parallel} dipole orientation reduces effectively to that 
for the power series
\begin{equation}
S_\parallel = \sum_{l=1}^\infty  
      l(l+1) \frac{\bar{\varepsilon} - 1}{\bar{\varepsilon}+ 1 + \frac{1}{l}}
       \left(\frac{a}{r_d}\right)^{2l}
=
 \frac{\bar{\varepsilon} - 1}{\bar{\varepsilon}+ 1} 
       \sum_{l=1}^\infty  \frac{l^2 (l+1)}{l + v} u^{l}
\label{spard}
\end{equation}
where $u$ and $v$ are the same as in eq \ref{uvd}.
In terms of $S_\parallel$,
\begin{equation}
\Gamma_{nr} (\parallel) = \frac{|\mbox{\boldmath $\mu$}|^2}{4\hbar\varepsilon_0 a^3} 
                       \, u^2 \, \mbox{Im } S_\parallel
\label{gmnrparps}
\end{equation}
When the iterative decomposition described above in connection to 
$S_\perp$ is applied to the coefficients 
of the power series for $S_\parallel$ in eq \ref{spard}, one finds
\begin{equation}
\frac{l^2(l+1)}{l+v} = 
   l^2 + l(1-v) - v (1-v) + \frac{v^2(1-v)}{l} 
                - \frac{v^3(1-v)}{l^2} + \frac{v^4(1-v)}{l^2(l+v)}
\label{opex2}
\end{equation}
Upon substituting eq \ref{opex2} back into eq \ref{spard}, and on using 
elementary summation formulas (eqs \ref{sf1}-\ref{sf4}), one arrives at
\begin{eqnarray}
\lefteqn{
u^2 \,S_\parallel = \frac{\bar{\varepsilon}- 1}{\bar{\varepsilon} + 1} \,u^2 \,
\left[
\frac{u(1+u)}{(1-u)^3} + \frac{u(1-v)}{(1-u)^2} 
\right.
}
\nonumber\\
&&
\left. 
   - \frac{u v (1-v)}{1-u}  - v^2(1-v)\, \ln(1-u) 
   - v^3(1-v)\, F (u,v)\right]
\label{spardf}
\end{eqnarray}
with the remainder $F(u,v)$ has been defined by eq \ref{fuv}.
Compared to eq \ref{sprdf}, the $u$-dependent terms in eq \ref{spardf} 
are, up to their prefactors, identical.
On substituting eq \ref{spardf} in eq \ref{gmnrparps}
\begin{eqnarray}
\lefteqn{
\Gamma_{nr} (\parallel) =  
  \frac{|\mbox{\boldmath $\mu$}|^2}{4\hbar\varepsilon_0 a^3} \, 
\mbox{Im }\left\{
\frac{\varepsilon - \varepsilon_0}{\varepsilon + \varepsilon_0} \,
\left[
\frac{u^3(1+u)}{(1-u)^3}  + \frac{u^3}{(1-u)^2} 
      \frac{\varepsilon}{\varepsilon+ \varepsilon_0}
-  \frac{u^3}{1-u}  
           \frac{\varepsilon_0\varepsilon}{(\varepsilon+ \varepsilon_0)^2}
\right.\right.
}
\nonumber\\
&&
\left.\left.
- \frac{\varepsilon_0^2 \varepsilon}{(\varepsilon+ \varepsilon_0)^3}\,  u^2\, \ln(1-u) 
- \frac{\varepsilon_0^3\varepsilon}{(\varepsilon + \varepsilon_0)^4}\, u^2\, F(u,v) \right]\right\}
\hspace*{2cm}
\label{gmnraf2}
\end{eqnarray}
where we have substituted for $v$ in eq \ref{spardf}
according to eq \ref{uvd}. 

The respective formulas (eqs \ref{gmnraf1} and \ref{gmnraf2}) provide 
the sought alternative representations of the quasistatic nonradiative 
rates Gersten and Nitzan \cite{GN,GNa} and Ford and Weber \cite{FoW1}
for the parallel and perpendicular dipole orientations.
No approximation has been used yet and the formulas of eqs \ref{gmnraf1} and \ref{gmnraf2}
are fully equivalent to the original GN expressions (eqs \ref{gmnrperp} and \ref{gmnrpar}).
The respective alternative representations comprise
four elementary analytic functions together with
a residual multipole series (cf eqs \ref{uvd}, \ref{fuv})
\begin{equation}
u^2\, F(u,v) = \sum_{l=1}^\infty 
    \frac{\varepsilon+\varepsilon_0}{l[l \varepsilon + (l+1)\varepsilon_0]} 
             \, u^{l+2}
\label{mms}
\end{equation} 
One can verify that the dependence of the nonradiative rates 
on $\varepsilon$ and $\varepsilon_0$ remains essentially only through 
the relative dielectric contrast 
$\bar{\varepsilon}$, whereas $\varepsilon_0$ merely multiplies 
a universal and particle size independent part expressed 
in terms of the relative parameters $u$ and $\bar{\varepsilon}$.
The elementary analytic functions in the square brackets 
in eqs \ref{gmnraf1} and \ref{gmnraf2} are ordered hierarchically 
from left to right according to  decreasing singularity 
in the short distance limit $d\rightarrow 0$.
Indeed, on making the use of dimensionless distance 
parameter $\delta=d/a\ll 1$ and on
expanding the binomial $u=[1+(d/a)]^{-2}=(1+\delta)^{-2}$,
one finds in the limit $\delta\rightarrow 0$
\begin{equation}
(1-u)^{-j} \sim (2\delta)^{-j}
\end{equation} 
The 4th term exhibits a logarithmic singularity $\sim \ln 2\delta$.
The power series $u^2 F$ has a finite 
limit for $\delta\rightarrow 0$, or equivalently $u\rightarrow 1$: the 
first series in the second equality in eq \ref{fuv} sums 
up to {\em dilogarithm} $\mbox{Li}_2(u)$ 
(see eq \ref{sf6}), which
is a {\em regular} function in the limit $u\rightarrow 1$ (see eq \ref{sf6f1}),
and the coefficients of the second series in the second equality 
in eq \ref{fuv} are for $l\gg 1$ bounded from above by those 
of $\mbox{Li}_2(u)$ (cf eq \ref{sf6}).

The reason why the respective iterative decompositions 
(eqs \ref{opex1} and \ref{opex2})
of the multipole contributions in the power series in eqs \ref{sprd} and \ref{spard} enable to disentangle various singular terms can 
easily be understood. The power series in eqs \ref{sprd} and \ref{spard} have 
the radius of convergence $|u|=1$, corresponding to $d=0$ (cf eq \ref{uvd}).
With the decomposition in eqs \ref{opex1} or \ref{opex2}
being effectively an expansion 
of the multipole contributions in eqs \ref{sprd} and \ref{spard}
into decreasing powers of $l$, each subsequent 
term in either eq \ref{opex1} or eq \ref{opex2} corresponds to 
the series contributing to a decreasing order of singularity 
in the limit $\delta\rightarrow 0$. Obviously, the described iterative decompositions 
could, in principle, be continued further also for the coefficients of the 
remainder $F(u,v)$ in eq \ref{fuv}.
The alternative representations (eqs \ref{gmnraf1} and \ref{gmnraf2})
of the nonradiative rates will be shown to

\begin{itemize}

\item possess a drastically improved convergence properties;

\item enable a compact analytic description of the GN nonradiative rates over
      entire parameter range;

\item enable one to derive a complete short-distance asymptotic behavior
      of the GN nonradiative rates in the dimensionless distance parameter $\delta$;
 
\item translate straightforwardly into corresponding results
        for the image enhancement factors $\Delta$.

\end{itemize}
The properties will be dealt with in the subsequent sections.

\section{Convergence properties}
\label{sc:cnv}
For practical calculations a cut-off $l_{max}$ has to be imposed 
on infinite power series in each of eqs \ref{gmnrperp}, \ref{gmnrpar}, 
and \ref{mms}. In what follows, the average nonradiative rates 
will be plotted. The latter are obtained
by angular averaging over different dipole orientations at a given dipole
radial position, resulting in (see eq 7 of ref \cite{Chew})
\begin{equation}
\bar{\Gamma} =[2\Gamma(\parallel) + \Gamma(\perp)]/3
\label{powav}
\end{equation}
(Note in passing that the rates $\Gamma(\parallel)$ 
enter with a factor $2$ into the orientational average,
because there are two linearly independent parallel 
dipole orientations possible.) The reason behind plotting the average
nonradiative rates is that they encompass the particular cases
of two relative dipole orientations considered in the previous section.
The average nonradiative rates obtained for a given cut-off $l_{max}$ 
will be plotted as the ratios to the converged nonradiative rates. 
The converged rates were here and below calculated with 
the cut-off $l_{max}=500$ in eqs 
\ref{gmnrperp} and \ref{gmnrpar}, in order to achieve
a sufficient convergence at an immediate sphere proximity.
Fortran F77 computer code is freely available on-line \cite{GNcd}.
It is well known a sphere may exhibit a series of
the $l$-polar localized surface plasmon resonances (LSPR), with
the  respective resonance conditions given by the formula \cite{BH}
\begin{equation}
\varepsilon=-\frac{l+1}{l}\,\varepsilon_0
\label{lsprc}
\end{equation}
Two particular cases are considered: (i) an {\em off-resonance case}, 
where the real part $\varepsilon'$ of $\varepsilon$ is 
well outside the resonance interval 
$- 2\varepsilon_0 \leq \varepsilon' < - \varepsilon_0$,
and (ii) an {\em on-resonance case}, 
where $\varepsilon'$ is close or within the resonance interval
while the imaginary part $\varepsilon''$ of $\varepsilon$ is relatively small.
For the sake of simplicity, the host 
medium will be in what follows assumed to be air ($\varepsilon_0=1$).

\subsection{Off-resonance case}
\label{sc:ofr}
Figure \ref{convr10l612} displays the so-called 
{\em average} nonradiative rates for the off-resonance case corresponding 
to a spherical nanoparticle with the radius $a=10$ nm, the emission wavelength 
of the dipole emitter $612$ nm, and the dielectric constant at 
the emission wavelength $\varepsilon(612 \mbox{ nm})= -15.04 + 1.02 i$ 
\cite{FoW1,AMoc}. 
The dependence of the nonradiative rates on the 
emission wavelength of a dipole
emitter is only {\em implicit} through $\varepsilon$ and $\varepsilon_0$.
Additionally, the dependence of the nonradiative rates
on the particle size only enters through the prefactor $a^{-3}$, which
multiplies a universal and particle size independent part expressed 
in terms of the dimensionless parameter $u$.
Therefore, although the above parameters were tailored to a silver 
nanoparticle (AgNP), the behavior shown in Figure \ref{convr10l612} is expected
to hold (at least qualitatively) for arbitrary 
particle sizes and sphere materials, as long
as one deals with an off-resonance case.
Typically, the situation in Figure \ref{convr10l612} corresponds 
to a rather common case of a fluorophore or a quantum dot 
with the emission wavelength of $\gtrsim 480$ nm 
in the case of a AgNP or with the emission wavelength of $\gtrsim 550$ nm 
in the case of a gold nanoparticle (AuNP).

The convergence of the conventional GN representations 
(eqs \ref{gmnrperp} and \ref{gmnrpar})
of nonradiative rates is illustrated in Figure \ref{convr10l612}a.
The conventional representations are characterized by a rapid
drop in precision at an immediate sphere proximity.
Even with $l_{max}=20$, the convergence for $d/a\lesssim 0.1$ 
is less than $77\%$.
Obviously, the nonradiative rates cannot be approximated by 
taking into account solely the dipole-dipole interaction \cite{Rup,AMoc}.
Compared to that, the alternative representations 
of nonradiative rates shown in Figure \ref{convr10l612}b 
exhibit a {\em superconvergence} - already for $l_{max}=1$, i.e.
by taking into account a mere residual dipole contribution in eq \ref{mms}, 
the alternative representations agree with the converged rates up to at least 
$99.95\%$ over entire distance interval.

\subsection{On-resonance case}
\label{sc:onr}
An {\em on-resonance} case is illustrated 
in Figure \ref{convr100l340}, which shows the ratios of the approximate
and converged average nonradiative rates for $a=100$, $\varepsilon_0=1$,
the emission wavelength of the dipole 
emitter $340$ nm, and $\varepsilon(340 \mbox{ nm})=-1 + 0.6i$. 
Similarly to  Figure \ref{convr10l612},
the latter value of $\varepsilon$ corresponds  to the dielectric constant
of AgNP at the wavelength slightly below $340$ nm. 
However, the situation shown in Figure \ref{convr100l340} 
is not generic, but rather specific to a metal with very low
imaginary part $\varepsilon''$ at the wavelength where $\varepsilon'=-1$.
For instance, in the case of Au one would find
$\varepsilon(400 \mbox{ nm})=-1.07 + 6.486 i$ \cite{Hand}, 
i.e. {\em ten-times} larger
$\varepsilon''$ than in the case of silver, which would rather correspond
to an off-resonance case (see also Section \ref{sc:at}).

Compared to the off-resonance case shown in 
Figure \ref{convr10l612}a,
the dipole contribution shown in Figure \ref{convr100l340}a 
is significantly reduced
at the corresponding values of $\delta$: from 
$\approx 50\%$ down to $\approx 30\%$ 
for $\delta=1$, and from $\approx 80\%$ down to $\approx 60\%$ for $\delta=2$. 
Compared to the off-resonance case of Figure \ref{convr10l612}b, 
the precision of the alternative representations shown 
in Figure \ref{convr100l340}b 
has for $l_{max}=1$ again a minimum at $\delta\approx 1$.
At $\delta\approx 1$ the precision drops from $99.95\%$ in the
off-resonance case shown in Figure \ref{convr10l612}b 
down to $87\%$ in the on-resonance case. The precision is still very good.
Note in passing that one customarily uses the respective $d^{-3}$ 
asymptotic behavior (see Figure \ref{gnnra10l612}) and 
$d^{-6}$ asymptotic behavior
(see the $l_{max}=1$ contributions in Figures \ref{convr10l612}a
and \ref{convr100l340}a) to fit experimental data even
if much lower precision (less than $80\%$) is guaranteed.
Interestingly, the inclusion of the residual quadrupole contributions for $l_{max}=2$ rapidly 
restores the convergence of the alternative representations, which becomes 
not worse than $98\%$ over entire length interval.
Therefore, the only difference compared to an off-resonance case is that
one should increase $l_{max}$ from $1$ to $2$
in order to obtain a fairly accurate description of the nonradiative rates.

\subsection{Summary of convergence properties}
\label{sc:scp}
It should be not surprising that the alternative representations have better
convergence properties. After all they were derived by summing exactly
the first four partial contributions of each given multipole in eqs
\ref{opex1} and \ref{opex2} up to an arbitrary high order. 
Since higher order multipoles 
in eqs \ref{gmnrperp} and \ref{gmnrpar} become increasingly relevant
with decreasing $d$, it was to be expected that 
the alternative representations would perform much better
in an immediate sphere proximity relative to eqs \ref{gmnrperp} and \ref{gmnrpar}.
What is surprizing here is an astonishingly fast convergence.

In order to understand the superconvergence, note the following. 
The coefficients of the power series (eq \ref{mms}), which takes into account 
residual multipole contributions,
are of the order ${\cal O}(l^{-2})$, whereas the coefficients of the original 
multipole series (eqs \ref{gmnrperp} and \ref{gmnrpar}) are of the order 
${\cal O}(l^{2})$ for $l\gg 1$. Therefore, beginning with 
the residual quadrupole ($l=2$) contribution, 
the residual contribution of higher order multipoles in the series (eq \ref{mms}) has been reduced by the factor of $l^{-4}$ compared 
to the original GN expressions (eqs \ref{gmnrperp} and \ref{gmnrpar}). 
Of course, the missing part of the multipole contributions 
did not disappear - it has become comprised in the 
analytic terms. The effect of higher-order multipoles comprised 
in the contribution 
of the elementary analytic functions in the square brackets 
in eqs \ref{gmnraf1} and \ref{gmnraf2} is appreciable over 
entire distance range. Indeed, the alternative 
representations shown in  Figure \ref{convr10l612}b agree 
with the converged rates within $0.03\%$ for 
$\delta\sim 2$ and $l_{max}=1$, 
whereas a mere $\sim 80\%$ precision is achieved by 
the dipolar term in the conventional GN representation 
(eqs \ref{gmnrperp} and \ref{gmnrpar}) of the nonradiative 
rates shown in Figure \ref{convr10l612}a.
Thus the drastically improved convergence of the alternative 
representations  (eqs \ref{gmnraf1}, \ref{gmnraf2}, \ref{mms}) 
can be viewed as a consequence of the fact that 
the elementary analytic functions of the alternative representations
(eqs \ref{gmnraf1}, \ref{gmnraf2}) comprise
an essential part of the contribution of 
an {\em infinite} number of higher order multipoles.
A slightly worse convergence properties in an on-resonance case is then
caused by an increase of the quadrupole 
contribution relative to the dipole one. The reason of that
increase will be further explained in Section \ref{sc:at}.

\section{The contribution of the analytic terms relative 
to that of the residual multipole series}
\label{sc:at}
In turns out expedient to determine of how much of the higher-order 
multipole contribution is comprised in each of the analytic terms 
in the square bracket of eqs \ref{gmnraf1} and \ref{gmnraf2}.
For that purpose we have considered various approximations 
to the average nonradiative rates, which were obtained 
by keeping a gradually increasing number of the analytic terms.
The results for an off-resonance case corresponding to
Figure \ref{convr10l612} are plotted in Figure \ref{gnnra10l612}.
It is obvious from Figure \ref{gnnra10l612} that
keeping mere first three terms in the 
square brackets in eqs \ref{gmnraf1} and \ref{gmnraf2} reproduces
the converged  rates already within $10\%$ over entire distance range,
whereas keeping the first four terms in the 
square brackets in eqs \ref{gmnraf1} and \ref{gmnraf2} reproduces
the converged rates within $1.2\%$ over entire distance range.
We recall that the last but one series in eq \ref{fuv}
sums to {\em dilogarithm} $\mbox{Li}_2(u)$. One can then substitute
\begin{equation}
F(u,v) = \mbox{Li}_2(u) - \varepsilon_0\, \sum_{l=1}^\infty 
    \frac{u^{l}}{l^2[l \varepsilon + (l+1)\varepsilon_0]}
\label{fuvd}
\end{equation}
into each of eqs \ref{gmnraf1} and \ref{gmnraf2}, 
and consider an approximation
which results by neglecting the last series, i.e. by approximating 
$F(u,v)$ by $\mbox{Li}_2(u)$.
Such an approximation [comprising all four singular terms together 
with $\mbox{Li}_2(u)$] then reproduces the 
exact results within remarkable $0.12\%$ 
over entire distance range. 
A similar agreement has also be obtained for $a=0.5$ nm AgNP as shown in 
Figure \ref{gnnrd1l612}.

The results shown in Figures \ref{gnnra10l612} and \ref{gnnrd1l612} suggest 
that our analytic terms have to comprise a substantial
part of the dipole contribution. Indeed, 
the dipole ($l_{max}=1$) contribution accounts for $\approx 50\%$ and 
$\approx 80\%$ of the average nonradiative rate for $\delta=1$ and $\delta=2$, 
respectively, in Figure \ref{convr10l612}a.
Therefore, without taking into account an essential part
of the  dipole contribution, less than $50\%$ agreement with the nonradiative
rates would be possible for $\delta\approx 1$ in an off-resonance case. 
By keeping the complete dipole contribution one could always achieve
a good agreement with the exact rates for sufficiently large $d$.
In the latter case the sphere response 
is that of a polarizable point, and the long-distance $d^{-6}$ dependence
is that corresponding to a {\em dipole-dipole} interaction \cite{FoW1}. 
Indeed, for $r_d\gg a$, so that one can approximate $r_d$ with $d$, 
the distance dependence of the nonradiative decay is obviously dominated by the 
$l=1$ dipole term in the GN series (eqs \ref{gmnrperp} and \ref{gmnrpar}), 
and the nonradiative decays follow a $d^{-6}$ distance dependence. 
Interestingly, the square brackets 
in eqs \ref{gmnraf1} and \ref{gmnraf2} exhibit a leading $d^{-6}$ ($\sim u^3$)
behavior in the long distance limit, in which case $u\ll 1$. 
Therefore, as long as the residual dipole contribution in the remainder 
$F(u,v)$ (defined by eq \ref{mms}) in eqs \ref{gmnraf1} 
and \ref{gmnraf2} is negligible compared 
to the original dipole contribution 
in eqs \ref{gmnrperp} and \ref{gmnrpar}, our analytical terms
are expected to provide very good approximation to the nonradiative rates.
Note in passing that the product 
$u^2\, F(u,v)$ in eqs \ref{gmnraf1} and \ref{gmnraf2} 
reduces for $l_{max}=1$ to (cf eq \ref{mms})
\begin{equation}
u^2\, F(u,v) \rightarrow \frac{\varepsilon+\varepsilon_0}{\varepsilon+2\varepsilon_0}\, u^3
\label{fuvl1}
\end{equation} 
Thus the condition that the residual dipole contribution in 
$F(u,v)$ is negligible compared to the original dipole 
contribution in eqs \ref{gmnrperp} and \ref{gmnrpar} translates into 
\begin{equation}
\left| \mbox{Im } 
\frac{c}{(\varepsilon + \varepsilon_0)^4}\, 
              \frac{\varepsilon - \varepsilon_0}
         {\varepsilon+ 2\varepsilon_0} \right| \ll
\left|\mbox{Im } \frac{\varepsilon - \varepsilon_0}
                        {\varepsilon+ 2\varepsilon_0}\right|
\label{cc}
\end{equation}
where $c=\varepsilon_0^2\varepsilon^2$ for a perpendicular dipole orientation and 
$c=\varepsilon_0^3\varepsilon$ for a parallel dipole orientation. 
Obviously, if the values of $\varepsilon$ and $\varepsilon_0$ are spaced apart
(e.g. $\varepsilon=-15.04 + 1.02 i$ and $\varepsilon_0=1$),
the left-hand side of eq \ref{cc} is strongly damped by the denominator of the first
fraction (e.g. by $\approx 14^2=196$ in the examples 
shown in Figures \ref{convr10l612}, \ref{gnnra10l612}, 
and \ref{gnnrd1l612}), and 
the condition (eq \ref{cc}) is satisfied. 

The condition in eq \ref{cc} enables one also to determine a worst 
case scenario, in which case the original multipole contributions
comprised in the analytic terms of our superconvergent 
representations (eqs \ref{gmnraf1} and \ref{gmnraf2}) are at a minimum.
The latter would occur for  $\varepsilon\approx - \varepsilon_0$, in which case 
the denominator of the first fraction on the left-hand side 
of eq \ref{cc} would work against the inequality \ref{cc}. 
Because the residual multipole contributions in the 
series (eq \ref{mms}) have been reduced 
by the factor of $l^{-4}$ compared 
to eqs \ref{gmnrperp} and \ref{gmnrpar}, a violation of 
the condition in eq \ref{cc} would not affect much high order multipoles.
Typically, only the relative residual contribution of a few low order multipoles
will be affected. This is exactly the scenario that has been 
illustrated in Figure \ref{convr100l340}, where 
it has been shown necessary to keep the residual quadrupole 
contribution in the remainder $F$ in order to 
ensure a reasonable convergence.

\section{Short-distance asymptotic behavior}
\label{sc:asp}
With decreasing $d$, higher order multipoles 
in eqs \ref{gmnrperp} and \ref{gmnrpar} become increasingly relevant.
Ultimately, in the limit of a sufficiently small dipole-sphere separation
$\delta=d/a\ll 1$, an {\em infinite} number of multipoles in eqs 
\ref{gmnrperp} and \ref{gmnrpar} contributes to 
the change of the $d^{-6}$ long distance dependence into the $d^{-3}$ 
short distance dependence (see eq B.24'' of ref \cite{GNa})
\begin{equation}
\Gamma_{nr}(\perp) \rightarrow 
 - \frac{|\mbox{\boldmath $\mu$}|^2 }{4\hbar \varepsilon_0 d^3} 
\, \mbox{Im }\frac{1}{\bar{\varepsilon}+ 1}
=
\frac{|\mbox{\boldmath $\mu$}|^2}{8\hbar \varepsilon_0 d^3} 
\, \mbox{Im }\frac{\bar{\varepsilon} - 1}{\bar{\varepsilon}+ 1}
\label{gmnrperpl}
\end{equation}
and (see eq B.45'' of ref \cite{GNa})
\begin{equation}
\Gamma_{nr}(\parallel) \rightarrow 
 - \frac{|\mbox{\boldmath $\mu$}|^2}{8 \hbar \varepsilon_0 d^3} 
\, \mbox{Im }\frac{1}{\bar{\varepsilon}+ 1}
=
 \frac{|\mbox{\boldmath $\mu$}|^2}{16 \hbar \varepsilon_0 d^3} 
\, \mbox{Im }\frac{\bar{\varepsilon} - 1}{\bar{\varepsilon}+ 1}
\label{gmnrparl}
\end{equation}
(The respective 2nd equalities in eqs \ref{gmnrperpl} and \ref{gmnrparl}  
have been obtained on using eq \ref{imid} in the limit $l\rightarrow \infty$.)
For a comparison, the power dissipated by an oscillating dipole 
at the distance $d$ above a {\em planar interface} of a metallic
half-space is (cf eq 3.23 of ref \cite{FoW1})
\begin{equation}
P_{abs}=\frac{\omega}{8 \varepsilon_0 d^3}\, 
   \left( \mbox{\boldmath $\mu$}_\perp^2 +\frac{1}{2}\mbox{\boldmath $\mu$}_\parallel^2\right)
    \mbox{Im } \frac{\varepsilon -\varepsilon_0}{\varepsilon +\varepsilon_0}
\label{fwdppl}
\end{equation}
One can easily verify that upon substituting $P_{abs}$ into 
the correspondence principle (eq \ref{qmcm}),
the limit values of nonradiative rates given by 
eqs \ref{gmnrperpl} and \ref{gmnrparl} are recovered. 
Thus the leading $d^{-3}$ short-distance dependence corresponds exactly to 
the case of a dipole located at the distance
$d$ above a half-space characterized by the dielectric constant $\varepsilon$.

Given the alternative representations 
(eqs \ref{gmnraf1} and \ref{gmnraf2}), it is possible 
to derive the short-distance asymptotic behavior of the 
Gersten-Nitzan expressions involving all singular terms 
in the limit $\delta\rightarrow 0$.
A slight nuisance is that the first term in the 
square bracket in eqs \ref{sprdf}
and \ref{spardf}
contributes also to the $\delta^{-2}$ and $\delta^{-1}$ 
terms (see eq \ref{o3}). Similarly, the second term in the 
square bracket in eqs \ref{sprdf} and \ref{spardf} contributes an 
additional $\delta^{-1}$ term 
(see eq \ref{o2}). 
After taking into account the sub-leading singular 
terms according to elementary formulas
(eqs \ref{o3}-\ref{o1} of Appendix \ref{sec:sf}) and substituting
back into eq \ref{sprdf}, one finds in the limit $\delta\rightarrow 0$
\begin{equation}
u^2\, S_\perp \sim \frac{a_{-3}}{\delta^{3}}  
         + \frac{a_{-2}}{\delta^{2}} + \frac{a_{-1}}{\delta} 
              + a_{log}\, \ln (2\delta) 
                      + {\cal O}(1)
\label{sprdsf}
\end{equation}
where
\begin{eqnarray}
a_{-3} &=& \frac{\varepsilon - \varepsilon_0}{4(\varepsilon + \varepsilon_0)},
\nonumber\\
a_{-2} &=& -
     \frac{(\varepsilon - \varepsilon_0)(\varepsilon+ 3 \varepsilon_0)}
        {8(\varepsilon+ \varepsilon_0)^2}
\nonumber\\
a_{-1} &=& 
    \frac{(\varepsilon - \varepsilon_0)(\varepsilon^2 + 3 \varepsilon_0^2)}
             {8(\varepsilon+ \varepsilon_0)^3}
\nonumber\\
a_{log} &=&
    \frac{\varepsilon_0\varepsilon^2(\varepsilon - \varepsilon_0)}
                {(\varepsilon+ \varepsilon_0)^4}
\label{amj}
\end{eqnarray}
On repeating the steps leading from eq \ref{sprdf} to eq \ref{sprdsf}, 
one finds in the limit $\delta\rightarrow 0$
\begin{equation}
u^2\, S_\parallel \sim  \frac{b_{-3}}{\delta^{3}}  
    + \frac{b_{-2}}{\delta^{2}} + \frac{b_{-1}}{\delta} 
        + b_{log}\, \ln (2\delta) 
            + {\cal O}(1),
\label{spardsf}
\end{equation}
where $b_{-3} \equiv a_{-3}$ and 
\begin{eqnarray}
b_{-2} &=& - 
     \frac{(\varepsilon - \varepsilon_0)(3\varepsilon+ 5 \varepsilon_0)}
           {8(\varepsilon+ \varepsilon_0)^2}
\nonumber\\
b_{-1} &=& 
   \frac{(\varepsilon - \varepsilon_0)
         (3\varepsilon^2+ 8\varepsilon_0\varepsilon + 9\varepsilon_0^2)}                   {8(\varepsilon+ \varepsilon_0)^3}
\nonumber\\
b_{log} &=& - 
    \frac{\varepsilon_0^2 \varepsilon (\varepsilon - \varepsilon_0)}
     {(\varepsilon+ \varepsilon_0)^4}
\label{bmj}
\end{eqnarray}
The corresponding short-distance asymptotic of the nonradiative rates follows then
straightforwardly on substituting eqs \ref{sprdsf} and \ref{spardsf} 
into eqs \ref{gmnrperps} and \ref{gmnrparps}, respectively.
As a consistency check, the term proportional to $\delta^{-3}$ 
could be shown to reproduce eqs \ref{gmnrperpl} and \ref{gmnrparl}, respectively.
Looking at the coefficients (cf eqs \ref{amj} and \ref{bmj})
of the asymptotic expansions, it is obvious that by a judicious choice 
of the dielectric constants $\varepsilon$ and $\varepsilon_0$ one could
switch off either $\delta^{-2}$ or $\delta^{-1}$ terms in 
eqs \ref{sprdsf} and \ref{spardsf}. For instance, 
\begin{eqnarray}
\varepsilon = - 3\varepsilon_0 &\Longrightarrow& a_{-2}=0
\nonumber\\
\varepsilon = -\frac{5}{3}\, \varepsilon_0 &\Longrightarrow& b_{-2}=0
\nonumber\\
\varepsilon = i \sqrt{3}\,\varepsilon_0 &\Longrightarrow& a_{-1}=0
\nonumber\\
\varepsilon = \left( -\frac{4}{3} + i\, \frac{\sqrt{11}}{3}\right)\, \varepsilon_0
 &\Longrightarrow& b_{-1}=0
\end{eqnarray}
A short-distance asymptotic behavior of the average nonradiative rates 
is shown in Figure \ref{asmpl340a100}. The 
configuration is the same  as in Figure \ref{convr100l340}.
Unfortunately, the use of the above asymptotic expansion is limited
to a rather short distance interval.
Indeed, the precision drops to below $90\%$ and $80\%$ for $\delta\approx 0.15$ 
and  $\delta\approx 0.2$, respectively. 
The major effect limiting the applicable distance range is
a limited range $\delta\lesssim 0.2$ of the validity of 
the asymptotic expansions (eqs \ref{o3}-\ref{o1}) of the 
elementary analytic functions in the square bracket 
in eqs \ref{gmnraf1} and \ref{gmnraf2} \cite{Coef}.
Note in passing that  the two-term asymptotic shows  
overall the best properties for an improved fitting, because it provides 
a significant improvement over the conventional $d^{-3}$ 
asymptotic over extended length scale. 
Keeping more terms in the asymptotic expansions (eqs \ref{sprdsf} and \ref{spardsf}) 
improves the asymptotic
in an immediate sphere proximity $\delta\ll 0.1$, but it comes at the expense
of worsening the precision for $\delta\gtrsim 0.25$.

\section{A comparison with the plane-surface 
$d^{-3}$ distance dependence} 
\label{sc:d3t}
In a large body of current literature one attempts to
fit experimental nonradiative rates with a power law dependence  $1/d^\sigma$,
where $\sigma=3$ for $\delta \lesssim 1$, and $\sigma$ is
between $3$ and $4$ when the separation satisfies $1\leq \delta \leq 4$
\cite{MTN,YJJ,JSS,PMS,HSP,BhN}. However, a $d^{-3}$ distance dependence 
(eqs \ref{gmnrperpl} and \ref{gmnrparl}) 
is typically limited to much smaller distance 
$d/a\lesssim 0.1$ (see Figure \ref{gnnra10l612}) and its precision rapidly decays
with the distance. Indeed, the leading $d^{-3}$ asymptotic equals {\em twice} the 
converged rates for $\delta \approx 0.5$
and {\em four-times} the converged rates for $\delta \approx 1$ (not shown) for the set-up 
considered in Figure \ref{asmpl340a100}.
If one wanted to remain within $10\%$ of the GN result, 
one is limited to $\delta \lesssim 0.1$,
which implies {\em sub-nanometer} distances for MNP radii $a\lesssim 10$ nm.
This is clearly not only unsatisfactory to reliably describe recent experiments 
\cite{DCL,MTN,Dulk,YJJ,JSS}, but also
the very use of the quasi-static rates is questionable for such small
distances. It is obvious from Figure \ref{asmpl340a100} that
a two term asymptotics yields a substantial
(essentially $100\%$) improvement in precision 
over the leading $d^{-3}$ short distance dependence: one 
can remain within $10\%$ of the GN result for $\delta \lesssim 0.18$ and
within $50\%$ of the GN result for $\delta \lesssim 0.5$.
The two-term asymptotic then equals {\em twice} the 
converged rates for $\delta \approx 1$ (not shown).
Another example of a substantial improvement of a leading asymptotic by 
a subleading term has been discussed in ref \cite{AMnw} (see Figure 4 therein).

In an intermediary region of distances $1\leq \delta \leq 4$
there appears to be no theoretical basis for any pure power law dependence 
such as $1/d^\sigma$, where $\sigma$ is between $3$ and $4$ 
\cite{MTN,YJJ,JSS,PMS,HSP}. Obviously, as shown in 
Figure \ref{asmpl340a100}, even a mixture of different negative 
powers of $\delta$ in full asymptotic expansions
(eqs. \ref{sprdsf} and \ref{spardsf})
yields rather poor approximation to the nonradiative rates. 
A significant improvement in describing the distance dependence 
of the nonradiative rates can instead be achieved by using the singular analytic
terms (see Figures \ref{gnnra10l612}-\ref{gnnrd1l612}).
In view of its simplicity, unmatched precision, and easy use
it is therefore preferable to use our superconvergent 
representations (eqs \ref{gmnraf1}, \ref{gmnraf2}, \ref{mms}) 
with a cut-off $l_{max}=1$, or in some cases $l_{max}=2$,
as fitting formulas for the experimental nonradiative rates.

\section{Image enhancement factors}
\label{sec:ief}
According to eq \ref{ief}, the value of $\Delta$ describes the change
in the net molecular dipole moment $\mbox{\boldmath $\mu$}$, which 
in turn determines the relevant
scattering cross sections. Therefore, the short distance behavior
of the respective image enhancement factors $\Delta$ is of interest 
for the surface-enhanced Raman scattering (SERS).
Given the respective definitions  
of $S_\perp$ and $S_\parallel$ (eqs \ref{sprd} and \ref{spard}), 
the image enhancement factors for the 
perpendicular and parallel dipole orientations are 
(after a trivial recasting of eq B.16' of ref \cite{GNa})
\begin{equation}
\Delta_\perp = \frac{\alpha}{a^3} \sum_{l=1}^\infty
 (l+1)^2\, \frac{\bar{\varepsilon}-1}{\bar{\varepsilon} +\frac{l+1}{l}} 
                    \, \left(\frac{a}{r_d}\right)^{2l+4}=\frac{\alpha}{a^3} 
                           \, u^2 \, S_\perp
\label{dtperp}
\end{equation}
and (see eq B.42' of ref \cite{GNa})
\begin{equation}
\Delta_\parallel = \frac{\alpha}{2 a^3} \sum_{l=1}^\infty
      l(l+1)\,\frac{\bar{\varepsilon}-1}{\bar{\varepsilon} + \frac{l+1}{l}} 
             \,\left(\frac{a}{r_d}\right)^{2l+4} 
                    = \frac{\alpha}{2a^3} \, u^2 \, S_\parallel
\label{dtpar}
\end{equation}
Therefore, the results obtained in preceding sections for the nonradiative
rates translate straightforwardly to those for the image enhancement factors
by substituting the relevant expressions for $u^2 \, S_\perp$ and 
$u^2 \, S_\parallel$ into eqs \ref{dtperp} and \ref{dtpar},
respectively.

\section{Outlook}
\label{sec:disc}
Our various approximation were extensively compared against
the converged {\em quasi-static} GN rates (eqs \ref{gmnrperp} and \ref{gmnrpar})
of Gersten and Nitzan \cite{GN,GNa} and Ford and Weber \cite{FoW1}.
The approximation could therefore be used whenever the 
Gersten and Nitzan theory \cite{GN,GNa} applies.
In some case one would probably be required to take into account the
effect of size corrections to the bulk dielectric function \cite{AMoc},
which can be straightforwardly incorporated.

We have only considered the nonradiative rates of fluorophores
and quantum dots relative to a MNP. However, recently a novel
class of composite superparticles have been introduced,
which possess so-called localized dielectric resonances (LDR)
having a quality factor comparable to that of LSPR \cite{AMsp}. 
The superparticles open a new avenue of applications for 
the radiative and nonradiative decay engineering, which will be followed
elsewhere.

On using a relation between polarizabilities and 
scattering T-matrix elements, one could contemplate
to include a dynamic depolarization and radiative reaction terms \cite{AMdp}
and consider a generalization of multipolar polarization factors according to
\begin{equation}
\frac{\bar{\varepsilon} - 1}{\bar{\varepsilon}+ \frac{l+1}{l}} 
\rightarrow \frac{(\bar{\varepsilon} - 1)
   \left[1-(\bar{\varepsilon}+1)\,\frac{x^2}{2(2l+3)}\right]}
{
\bar{\varepsilon} + \frac{l+1}{l}
+\frac{x^2}{2l(2l+3)}\left(-l\bar{\varepsilon}^2 
       -\frac{3(2l+1)}{2l-1}\,\bar{\varepsilon}
+\frac{(l+1)(2l+3)}{2l-1}\right)
- i\frac{(l+1)x^{2l+1}}{l(2l-1)!!(2l+1)!!}\, (\bar{\varepsilon} - 1)
}
\label{pleq}
\end{equation}
where $x=k_0a$ is the conventional size parameter.
Such a generalization is irrelevant for small particles \cite{AMoc}, but
it would be of interest 
for larger particles ($a\gtrsim 80$ nm), where depolarization ($\sim x^2$ term in 
eq \ref{pleq})
and radiative reaction become increasingly important \cite{AMdp}.
However, to implement it in all orders, in order to obtain a generalization
of our superconvergent representations, may be difficult. 
Nevertheless, it turns that an analogous
generalization of mere dipole order could already provide a sizable improvement
for larger particles ($a\gtrsim 80$ nm) \cite{MKP} and may sufficiently agree
with exact electrodynamic calculations \cite{Rup,KLG2,MKP,Ch,Chew}.
In this regard note that the results of Merten et al \cite{MKP} could be
improved by using a proper dynamic depolarization term 
that yields a correct dipole LSPR position 
up to the $x^2$-order \cite{AMdp}.

The nonradiative rates are often plotted 
normalized with respect to the radiative 
rates of a dipole in an infinite host medium. 
On substituting the Larmor formula for the total dipole
radiative power
\begin{equation}
P_r =\frac{\omega^4 |\mbox{\boldmath $\mu$}|^2}{3c^3}\, \sqrt{\varepsilon_0}
\end{equation}
into the correspondence principle (eq \ref{qmcm}), the 
dipole radiative rate in the host medium in the 
sphere absence is
\begin{equation}
\Gamma_r =\frac{\omega^3 |\mbox{\boldmath $\mu$}|^2}{3\hbar c^3}\, 
  \sqrt{\varepsilon_0}.
\end{equation}
Therefore, the normalization has the net effect of 
replacing the prefactor in the absolute
nonradiative rates (eqs \ref{gmnrperp}, \ref{gmnrpar}, 
\ref{gmnraf1}, \ref{gmnraf2})  
according to
\begin{equation}
\frac{|\mbox{\boldmath $\mu$}|^2}{\hbar\varepsilon_0 a^3} 
\rightarrow \frac{3}{x^3}
\end{equation}

\section{Conclusions}
\label{sec:conc}
In a large body of current literature one attempts to
fit experimental nonradiative rates with a power law dependence  $1/d^\sigma$,
where $\sigma=3$ for $\delta \lesssim 1$, and $\sigma$ is
between $3$ and $4$ when the separation satisfies $1\leq \delta \leq 4$
\cite{MTN,YJJ,JSS,PMS,HSP}. We have shown that such an approximation
could be highly imprecise
(see Figures \ref{gnnra10l612}, \ref{asmpl340a100}b). Instead, alternative superconvergent representations (eqs \ref{gmnraf1}, \ref{gmnraf2}, \ref{mms}) 
of the quasistatic nonradiative rates of Gersten and Nitzan \cite{GN,GNa} 
and Ford and Weber \cite{FoW1} were derived, which could be used for highly
precise, simple, and efficient analytic description of the rates.
Given the distance $d$ of a dipole from a sphere surface of radius $a$,
the representations comprise 
four elementary analytic functions and a modified series (eq \ref{mms}) taking into account 
residual multipole contributions. 
The analytic functions could be arranged hierarchically according
to decreasing singularity at the short 
distance limit $d\rightarrow 0$, ranging from $d^{-3}$ over $d^{-1}$ to $\ln (d/a)$. 
In the opposite long distance limit, the analytic functions exhibit a leading $d^{-6}$
behavior. On keeping mere residual dipole contribution of the infinite 
series (eq \ref{mms}), the  
representations typically agree with the converged rates over at least   
$99.9\%$ over entire length intervals, 
for arbitrary particle sizes and emission wavelengths, and 
for a broad range of dielectric constants (see Figure \ref{convr10l612}b).
The origin of the superconvergence was identified and explained.
The analytic terms of the representations reveal a complex distance dependence and
could be used to smoothly interpolate between the familiar $d^{-3}$ 
short-distance and $d^{-6}$ long-distance behaviors with an unprecedented accuracy
(see Figures \ref{gnnra10l612}, \ref{gnnrd1l612}). 
Therefore, the representations could be especially useful for 
the qualitative and quantitative understanding 
of the distance behavior of nonradiative rates 
of fluorophores and semiconductor
quantum dots involving nanometal surface energy transfer in the presence 
of metallic nanoparticles or nanoantennas.
As a byproduct, a complete short-distance asymptotic 
of the quasistatic nonradiative rates was derived.
The above results for the nonradiative rates translate
straightforwardly to the so-called image enhancement factors $\Delta$,
which are of relevance for the surface-enhanced Raman scattering (SERS).

The radiative rate in the GN theory is given by a single dipole term 
(cf eqs B.18' and B.43' of ref \cite{GNa}). The total rate in the GN theory 
is then obtained by adding the dipole radiative rate term to the nonradiative rate.
Therefore our results extend straightfowradly also to the total rate 
in the GN theory.

\section{Supporting Information Available:}
You can download the data of the plots and view the plots in different scales by editing 
options of the corresponding Origin${}^{\circledR}$ projects available on-line 
at http://www.wave-scattering.com/gn.html.

\newpage

\appendix

\section{Elementary summation and expansion formulas}
\label{sec:sf}
\begin{equation}
\sum_{l=1}\, u^l =\frac{u}{1-u}
\label{sf1}
\end{equation}
\begin{equation}
\sum_{l=1}\, l\, u^l =\frac{u}{(1-u)^2}
\label{sf2}
\end{equation}
\begin{equation}
\sum_{l=1}\, l^2\, u^l =\frac{u(1+u)}{(1-u)^3}
\label{sf3}
\end{equation} 
\begin{equation}
\sum_{l=1}\, \frac{u^l}{l} = - \ln (1-u)
\label{sf4}
\end{equation}
and (see eq 5.2.5.4 of ref \cite{PBM})
\begin{equation}
\sum_{l=1}\, \frac{u^l}{l^2} = - \int_0^u \frac{\ln (1-s)}{s}\, ds =
\mbox{Li}_2(u)
\label{sf6}
\end{equation}
defines the so-called {\em dilogarithm} function. 
The identity in eq \ref{sf3} follows from (see eq 5.2.2.9 of ref \cite{PBM})
\begin{equation}
\sum_{l=0}\, (l+1)^2\, x^l =\frac{1+x}{(1-x)^3}
\label{els2}
\end{equation}
The identities in eqs \ref{sf2} and \ref{sf3} can also be obtained on 
successively applying the operator $[u(d/du)]$ on both sides of eq \ref{sf1}.
The dilogarithm value at the limiting case of $u=1$ 
is {\em finite} and related to the Riemann zeta function
\begin{equation}
\mbox{Li}_2(1)=\zeta(2)=\frac{\pi^2}{6}
\label{sf6f1}
\end{equation}

In the limit $\delta\rightarrow 0$ one finds upon invoking the binomial series
\begin{eqnarray}
u   &=& (1+\delta)^{-2} \sim 1 - 2\delta + 3\delta^2 
   - 4\delta^3 + 5\delta^4 - 6\delta^5 = 1 - 2\delta (1-q)
\nonumber\\
u^2 &=& (1+\delta)^{-4} \sim 1 - 4\delta + 10 \delta^2 - 20 \delta^3 + 35 \delta^4
\nonumber\\
u^3 &=& (1+\delta)^{-6} \sim 1 - 6\delta + 21 \delta^2 - 56 \delta^3 + 126 \delta^4
\nonumber\\
u^4 &=& (1+\delta)^{-8} \sim 1 - 8\delta + 36 \delta^2 - 120 \delta^3 + 330 \delta^4
\nonumber\\
u^3(1+u) &\sim& 2 - 14\delta + 57 \delta^2 - 176 \delta^3 + 456 \delta^4
\nonumber\\
(1-u)^{-1} &\sim&  \frac{1}{2\delta}\, 
        \left(1 + \frac{3}{2}\, \delta + \frac{1}{4}\, \delta^2 - \frac{1}{8}\, \delta^3 \right)
\nonumber\\
(1-u)^{-2} &\sim&  \frac{1}{4\delta^2}\, 
                \left(1 + 3\delta + \frac{11}{4}\, \delta^2 + \frac{1}{2}\, \delta^3 \right)
\nonumber\\
(1-u)^{-3} &\sim&  
      \frac{1}{8\delta^3}\, \left(1 + \frac{9}{2}\, \delta 
                + \frac{15}{2}\, \delta^2 + \frac{21}{4}\, \delta^3 \right)
\nonumber
\end{eqnarray}
and
\begin{eqnarray}
\frac{u^3(1+u)}{(1-u)^{3}} &\sim&
           \frac{1}{4\delta^3} - \frac{5}{8\delta^2} + \frac{9}{8\delta} + {\cal O}(1)
\label{o3}
\\
\frac{u^3}{(1-u)^{2}} &\sim& \frac{1}{4\delta^2} - \frac{3}{4\delta}  + \frac{23}{16} 
                + {\cal O}(\delta^2)
\label{o2}
\\
\frac{u^3}{1-u}  &\sim&        \frac{1}{2\delta} -\frac{9}{4} + 
                    \frac{49}{8} \, \delta - 
                              \frac{209}{16} \, \delta^2 + {\cal O}(\delta^3)
\label{o1}
\end{eqnarray}


\newpage


\begin{center}
{\large\bf Figure captions}
\end{center}

\vspace*{0.4cm}

\noindent {\bf Figure 1 -} 
An illustration of the geometry of the problem.
Sphere of radius $a$ is located at the
coordinate origin. The dipole is positioned outside the
sphere at the distance $r_d=|{\bf r}_d|>a$ from the sphere center.

\vspace*{0.4cm}

\noindent {\bf Figure 2 -}
Convergence properties of the conventional 
GN representations (eqs \ref{gmnrperp}, \ref{gmnrpar})
of the nonradiative rates {(\bf a)}
and novel superconvergent representations 
(eqs \ref{gmnraf1}, \ref{gmnraf2}, \ref{mms})
of the nonradiative rates {(\bf b)}.
The average nonradiative rates (eq \ref{powav}) were calculated 
for the case of a spherical nanoparticle
with radius $a=10$ nm in air ($\varepsilon_0=1$), the emission 
wavelength of the dipole 
emitter $612$ nm, and the dielectric constant at 
the emission wavelength 
$\varepsilon(612 \mbox{ nm})= -15.04 + 1.02 i$ \cite{FoW1,AMoc}.
The parameters correspond to a silver spherical 
nanoparticle (AgNP), but an analogous
behaviour is expected in any {\em off-resonance} case. 
The nonradiative rates are plotted as the ratios 
of the approximate
to converged rates. The latter were 
calculated with the cut-off $l_{max}=500$ in eqs 
\ref{gmnrperp}, \ref{gmnrpar} 
in order to achieve a sufficient convergence at a sphere proximity.
Note different scale on the ordinate axis for 
{(\bf a)} and {(\bf b)}.

\vspace*{0.4cm}

\noindent {\bf Figure 3 -}
Convergence properties of the conventional 
representations (eqs \ref{gmnrperp}, \ref{gmnrpar})
of the nonradiative rates {(\bf a)}
and novel representations (eqs \ref{gmnraf1}), \ref{gmnraf2}), \ref{mms})
of the nonradiative rates {(\bf b)} in an {\em on-resonance} case.
The average nonradiative rates (eq \ref{powav}) were calculated 
for the case of a silver spherical nanoparticle (AgNP) 
with radius $a=100$ nm and the emission wavelength of the dipole 
emitter $340$ nm. The dielectric constant of AgNP at 
the emission wavelength was taken to be 
$\varepsilon(340 \mbox{ nm})= -1 + 0.6 i$.
Note different scale on the ordinate axis for 
{(\bf a)} and {(\bf b)}.

\vspace*{0.4cm}

\begin{figure}[tbp]
\begin{center}
\epsfig{file=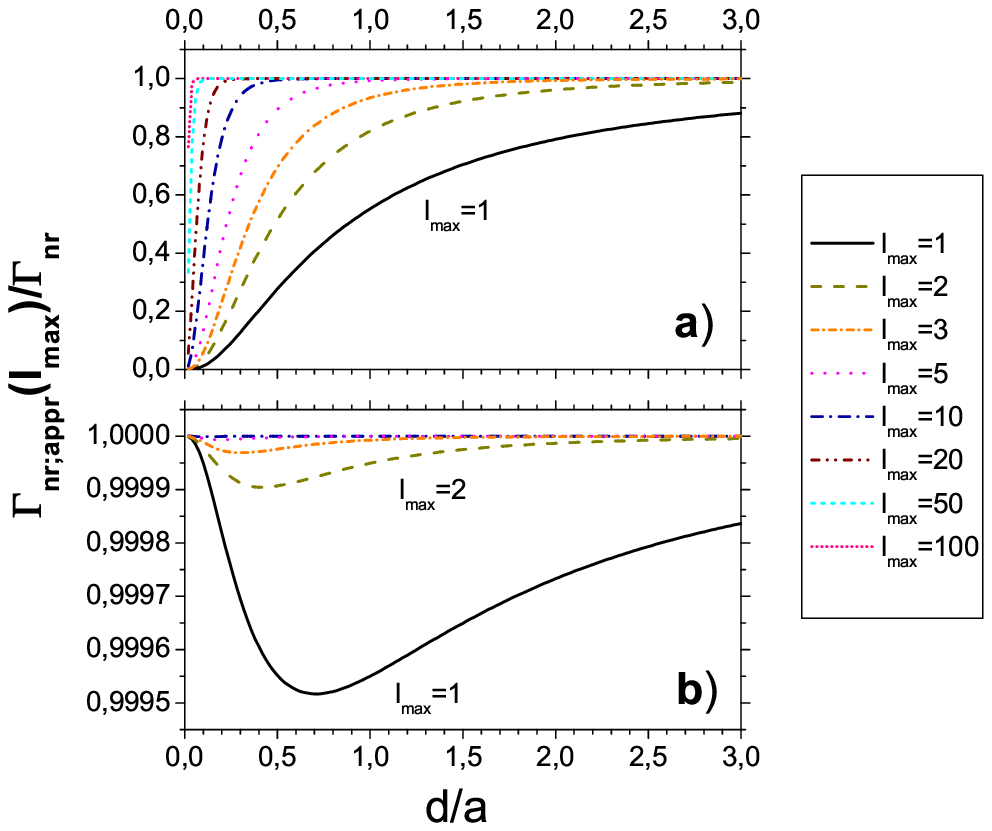,width=14cm,clip=0,angle=0}
\end{center}
\caption{}
\label{convr10l612}
\end{figure}

\noindent {\bf Figure 4 -} 
Comparison of various approximations to the orientationally 
averaged nonradiative rates (eq \ref{powav}) obtained by
keeping a gradually increasing number of 
analytic terms from left to right in the square bracket 
in eqs \ref{gmnraf1} and \ref{gmnraf2}. 
The average nonradiative rates are plotted as the ratios of approximate
to converged rates for the case of AgNP with radius $a=10$ nm in air 
($\varepsilon_0=1$), the emission wavelength of the dipole emitter $612$ nm, and 
$\varepsilon(612 \mbox{ nm})= -15.04 + 1.02 i$ 
(the same configuration as in Figure \ref{convr10l612}). 
The plane-surface $d^{-3}$ distance dependence  
is also shown for a comparison.
Keeping the first four terms in the 
square bracket in eqs \ref{gmnraf1} and \ref{gmnraf2} reproduces
the exact results within $1.2\%$ over entire distance range.
Upon approximating $F(u,v)$ by the dilogarithm term $\mbox{Li}_2(u)$ 
(eq \ref{fuvd}), the exact results are reproduced by resulting
five analytic terms within $0.12\%$ 
over entire distance range. The last two approximations 
essentially overlay into a single line.

\vspace*{0.4cm}

\noindent {\bf Figure 5 -}
The same as in Figure \ref{gnnra10l612} but for a AgNP of radius $a=0.5$ nm
($\varepsilon_0=1$, the emission wavelength $612$ nm, and $\varepsilon(612 \mbox{ nm})= -15.04 + 1.02 i$).

\vspace*{0.4cm}

\begin{figure}[tbp]
\begin{center}
\epsfig{file=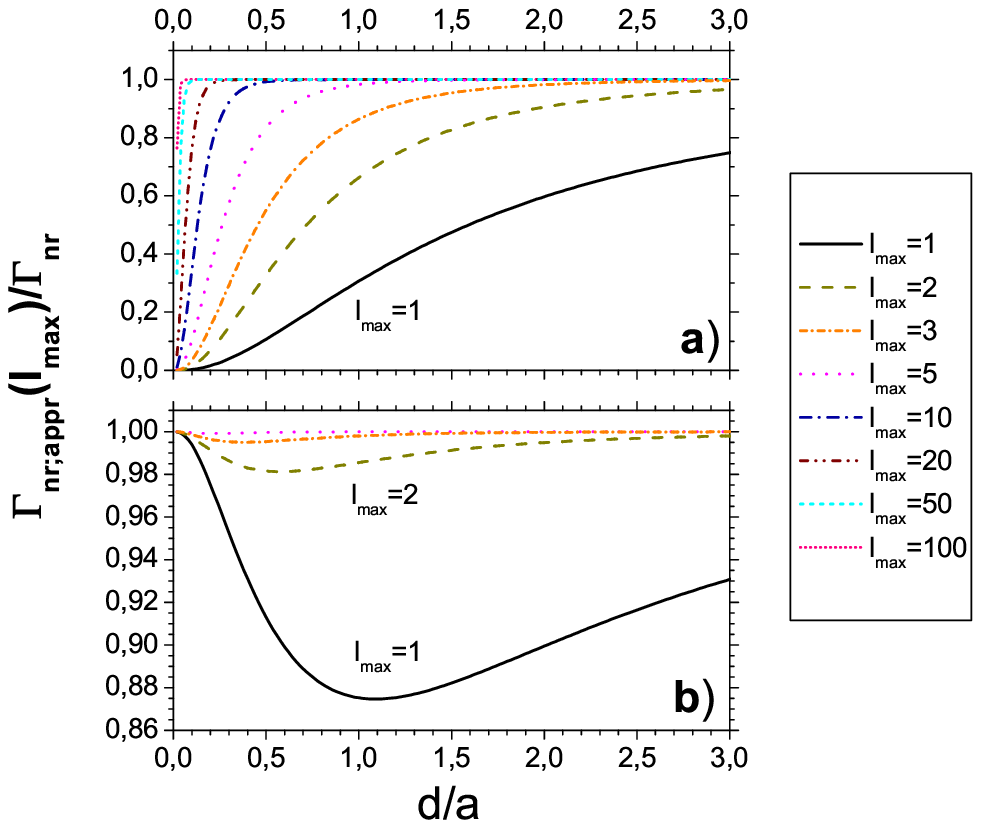,width=14cm,clip=0,angle=0}
\end{center}
\caption{}
\label{convr100l340}
\end{figure}

\begin{figure}[tbp]
\begin{center}
\epsfig{file=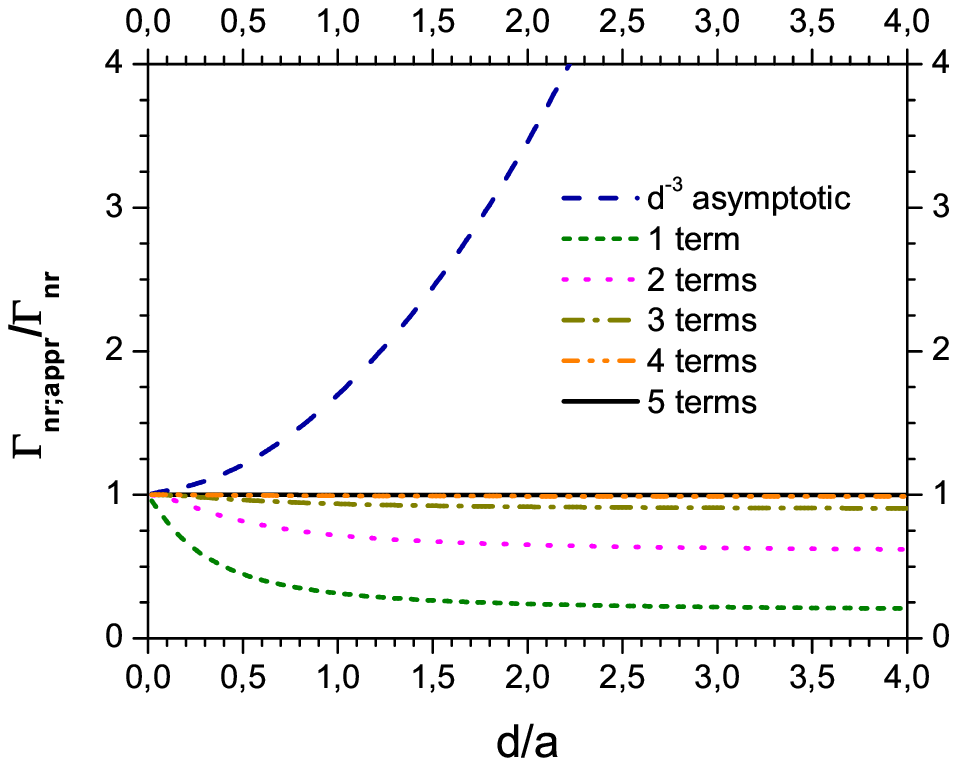,width=14cm,clip=0,angle=0}
\end{center}
\caption{}
\label{gnnra10l612}
\end{figure}

\begin{figure}[tbp]
\begin{center}
\epsfig{file=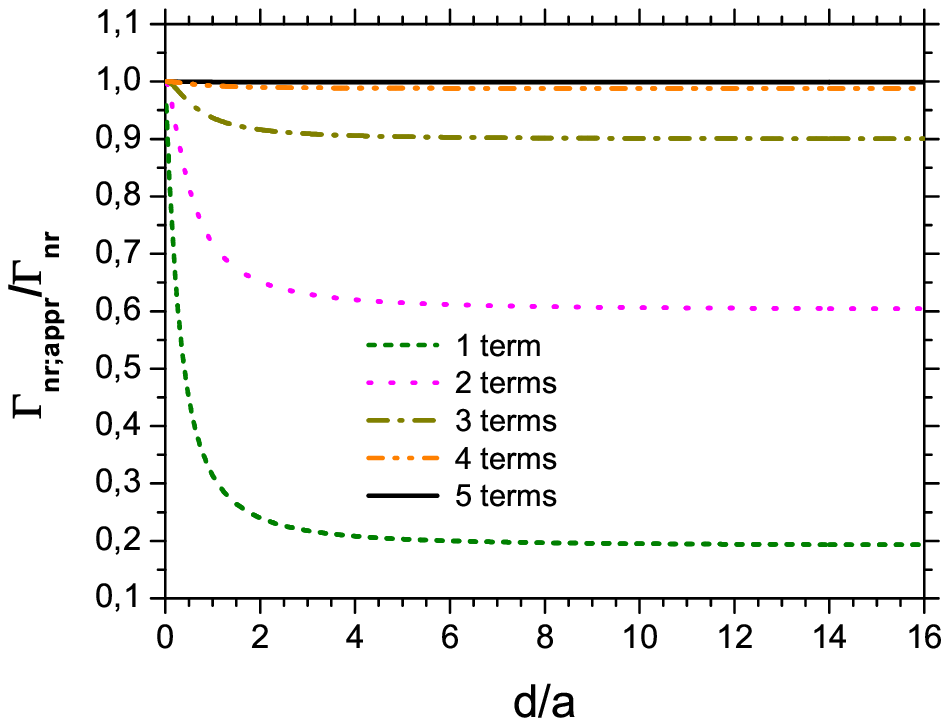,width=14cm,clip=0,angle=0}
\end{center}
\caption{}
\label{gnnrd1l612}
\end{figure}

\begin{figure}[tbp]
\begin{center}
\epsfig{file=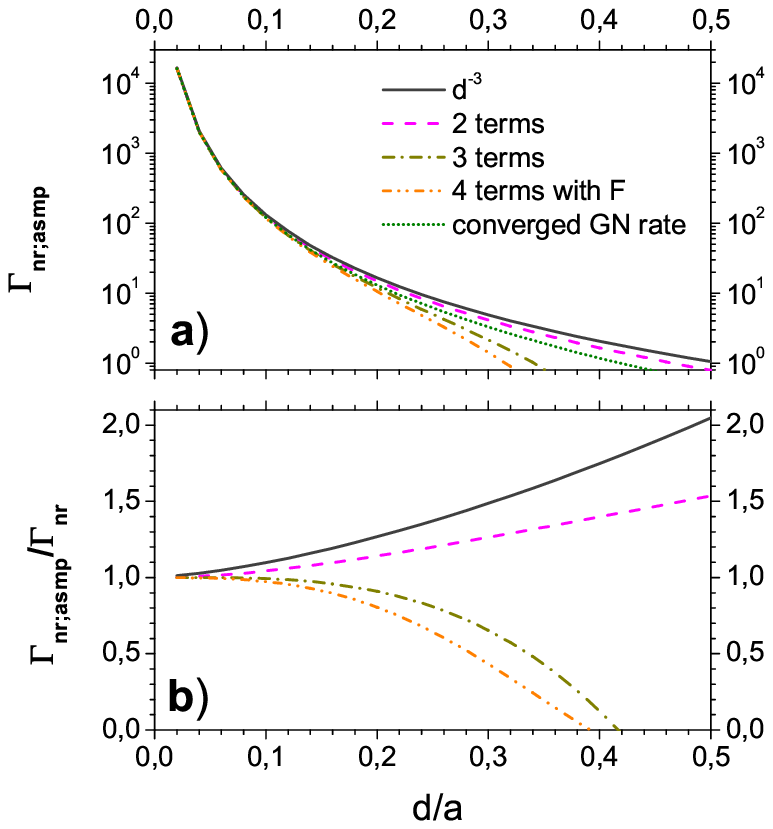,width=14cm,clip=0,angle=0}
\end{center}
\caption{}
\label{asmpl340a100}
\end{figure}

\noindent {\bf Figure 6 -}
A short-distance asymptotic behavior of the average nonradiative rates for
the case of AgNP with radius $a=100$ nm in air, the emission wavelength of the dipole 
emitter $340$ nm, and $\varepsilon(340 \mbox{ nm})= -1 + 0.6 i$ 
(the same configuration as in Figure \ref{convr100l340}). The Figure displays
various approximations obtained by 
keeping an increasing number of terms in eqs \ref{sprdsf} and \ref{spardsf}.
The converged GN rates are also shown for a comparison.
Obviously plotting in the logarithmic scale hides to a large extent 
the differences with the converged GN rates.
A comparision of the analytic expressions in eqs \ref{gmnraf1} and \ref{gmnraf2} 
with the inverse cube $d^{-3}$ plane-surface distance dependence 
has been shown in Figure \ref{gnnra10l612}.

\newpage

\newpage

\newpage

\newpage

\newpage

\newpage

\newpage

\begin{thebibliography}{99}



\bibitem{DCL} 
Dubertret, B.; Calame, M.; Libchaber, A. J.
Single-mismatch detection using gold-quenched fluorescent 
oligonucleotides.
Nat. Biotechnol. {\bf 2001}, 19, 365-370.


\bibitem{MTN} 
Maxwell, D. J.; Taylor, J. R.; Nie, S. 
Self-assembled nanoparticle probes for recognition 
and detection of biomolecules.
J. Am. Chem. Soc. {\bf 2002}, 124, 9606-9612.


\bibitem{Dulk}
Dulkeith, E.; Morteani, A. C.; Niedereichholz, T.; Klar, T. A.; Feldmann, J.;
 Levi, S. A.; van Veggel, F. C. J. M.; Reinhoudt, D. N.; M\"{o}ller, M.; 
Gittins, D. I. 
Fluorescence quenching of dye
molecules near gold nanoparticles: Radiative and  nonradiative effects.
Phys. Rev. Lett. {\bf 2002}, 89, 203002.


\bibitem{Lak5}
Lakowicz, J. R. 
Radiative decay engineering 5: metal-enhanced 
fluorescence and plasmon emission.
Anal. Biochem. {\bf 2005}, 337, 171-194.


\bibitem{YJJ}
Yun, C. S.; Javier, A.; Jennings, T.; Fisher, M.; Hira, S.; Peterson, S.;
 Hopkins, B.; Reich, N. O.; Strouse, G. F.
Nanometal surface energy transfer in optical rulers, 
breaking the FRET barrier.
J. Am. Chem. Soc. {\bf 2005}, 127, 3115-3119.


\bibitem{JSS}
Jennings, T. L.; Schlatterer, J. C.; Singh, M. P.; Greenbaum, N. L.;
 Strouse, G. F.
NSET molecular beacon analysis of hammerhead RNA substrate binding 
and catalysis.
Nano Lett. {\bf 2006}, 6, 1318-1324.


\bibitem{SRW}
Soller, T.; Ringler, M.; Wunderlich, M.; Klar, T. A.; 
Feldmann, J.; Josel, H. P.; Markert, Y.; Nichtl, A.; K\"{u}rzinger, K. 
Radiative and nonradiative rates of phosphors attached to 
gold nanoparticles. 
Nano Lett. {\bf 2007}, 7, 1941-1946.


\bibitem{PMS}
Pons, T.; Medintz, I. L.; Sapsford, K. E.; Higashiya, S.; 
Grimes, A. F.; English, D. S.; Mattoussi, H. 
On the quenching of semiconductor quantum dot photoluminescence 
by proximal gold nanoparticles. 
Nano Lett. {\bf 2007}, 7, 3157-3164.


\bibitem{HSP} 
Haldar, K. K.; Sen, T.; Patra, A.
Metal Conjugated Semiconductor Hybrid Nanoparticle-Based 
Fluorescence Resonance Energy Transfer.
J. Phys. Chem. C {\bf 2010}, 114, pp 4869-4874.


\bibitem{Rup}
Ruppin, R.
 Decay of an excited molecule near a small metal sphere.
J. Chem. Phys. {\bf 1982}, 76, 1681-1684.


\bibitem{KLG2}
Kim, Y. S.; Leung, P. T.; George, T. F.
Classical decay rates for molecules in the
presence of a  spherical surface: A complete treatment. 
Surf. Sci. {\bf 1988}, 195, 1-14.


\bibitem{GN}
Gersten, J.; Nitzan, A.
Spectroscopic properties of molecules interacting with
small dielectric particles. 
J. Chem. Phys. {\bf 1981}, 75, 1139-1152.


\bibitem{FoW1}
Ford, G. W.; Weber, W. H. 
Electromagnetic interactions of molecules with metal surfaces.
Phys. Rep. {\bf 1984}, 113, 195-287.


\bibitem{AMap}
Moroz, A. 
A recursive transfer-matrix solution 
for a dipole radiating inside and outside a stratified sphere.
 Ann. Phys. (NY) {\bf 2005}, 315, 352-418.


\bibitem{AMcl}
Moroz, A. Spectroscopic properties of a two-level atom 
interacting with a complex spherical nanoshell.
Chem. Phys. {\bf 2005}, 317, 1-15.


\bibitem{MKP}
Mertens, H.; Koenderink, A. F.; Polman, A. 
Plasmon-enhanced luminescence near noble-metal nanospheres: Comparison of 
exact theory and an improved Gersten and Nitzan model.
Phys. Rev. B {\bf 2007}, 76, 115123.


\bibitem{AMoc}
Moroz, A. 
Non-radiative decay of a dipole emitter 
close to a metallic nanoparticle: 
Importance of higher-order multipole contributions.
Opt. Commun. {\bf 2010}, 283, 2277-2287.


\bibitem{Ch}
Chew, H.
Transition rates of atoms near spherical surfaces.
J. Chem. Phys. {\bf 1987}, 87, 1355-1360.


\bibitem{Chew}
Chew, H. 
Radiation and lifetimes of atoms inside dielectric particles.
Phys. Rev. A {\bf 1988}, 38, 3410-3416.


\bibitem{AMcd}
The source codes CHEWFS and CHEW could be downloaded from 
http://www.wave-scattering.com/chew.f and 
http://www.wave-scattering.com/chewfs.f. 
A brief description of the codes can be downloaded from 
http://www.wave-scattering.com/chew-man.pdf


\bibitem{GNa}
Mathematical appendices for \cite{GN} (AIP Document No. PAPS JCP
SA-75-1139-32) available 
as on-line supplementary material 
(see also \verb|http://atto.tau.ac.il/~nitzan/68-appendix.pdf|).


\bibitem{GNcd}
F77 source code based on Gersten and Nitzan theory \cite{GN,GNa}, which
has been used to generate results here, could be downloaded from 
http://www.wave-scattering.com/gn.f.


\bibitem{BH}
Bohren C. F.; Huffman, D. R. {\em Absorption and 
Scattering of Light by Small Particles}, 
John Wiley \& Sons, New York, {\bf 1988}. 



\bibitem{Hand}
Palik, E. D., ed., 
{\em Handbook of Optical Constants of Solids}; Academic
Press: New York, 1985.


\bibitem{Coef}
A comparison of the respective analytic terms with their 
short-distance asymptotic could be downloaded from 
http://www.wave-scattering.com/Coefasmp.txt. 


\bibitem{BhN}
Bharadwaj, P.; Novotny, L. 
Spectral dependence of single molecule 
fluorescence enhancement.
Opt. Express {\bf 2007}, 15, 14266-14274. 
 

\bibitem{AMnw}
Moroz, A.
Electron mean-free path in metal coated
nanowires. J. Opt. Soc. Am. B {\bf 2011}, 28, 1130-1138.


\bibitem{AMsp}
Moroz, A. Localized resonances of composite particles.
  J. Phys. Chem. C {\bf 2009}, 113, 21604-21610.


\bibitem{AMdp}
Moroz, A.
Depolarization field of spheroidal particles.
J. Opt. Soc. Am. B {\bf 2009}, 26, 517-527. 
 

\bibitem{PBM}
Prudnikov, A. P.; Brychkov; Yu, A.; Marichev, O. I. 
{\em Integrals and Series}, 
2nd ed; Gordon and Breach: London, {\bf 1988}.



\end{thebibliography}
\end{document}